\documentclass[a4paper, 12pt]{article}
\usepackage{graphicx}
\usepackage{amsmath}
\usepackage{amssymb}

\begin{document}

\title{The filamentation of the laser beam as a \emph{labyrinth} instability}
\author{F. Spineanu and M. Vlad \\
National Institute of Laser, Plasma and Radiation Physics\\
Magurele, Bucharest 077125, Romania\\
{\it florin.spineanu@free.fr}}
\date{}
\maketitle

\begin{abstract}
At incident powers much higher than the threshold for filamentation a pulse from a high-power laser generates in the transversal plane a complex structure. It consists of randomly meandering stripes defining connected regions where the field intensity is high; and, the complementary regions dominated by diffusive plasma with defocusing property. The pattern is similar to an ensemble of clusters of various extensions. 
    We provide evidence that there is a correlation between this filamentation and the {\it labyrinth} instability in reaction-diffusion systems. Besides the similarity of the spatial organization in the two cases, we show that the differential equations that describe these two dynamical processes lead to effects that can be mutually mapped. For the laser beam at high power the Non-linear Schrodinger Equation in a regime of strong self-focusing and ionization of the air leads to multiple filamentation and the structure of clusters. Under the effect of the {\it labyrinth} instability a model of activator-inhibitor leads to a similar pattern. The origin of this connection must be found in the fact that both optical turbulence and the activator-inhibitor dynamics have the nature of competition between two phases of the same system.
\end{abstract}

\section{Introduction}

The propagation of the beam of a high power laser is an interaction with the
atmospheric medium that involves the local activation of the polarization
and of other processes that depends nonlinearly on the incident field. The
first consequence is the tendency of self-suppression of the propagation,
caused by the self-focusing of the beam. In an analytic description this is
manifested by the existance of a singularity of the electric field of the
wave, which arises in finite time \cite{Marburger1975}. Effectively, the characteristic
trajectories of the propagation converge toward a caustic. The propagation
(as a beam) has no meaning beyond this point.

In reality, the dynamics is more complex, including wave diffraction and
dispersion of the group velocity, which act to prevent the singularity \cite{Couaironfilamentation}, \cite{bergephysrep}. More
important, the self-focusing increases the intensity ($\sim \left\vert 
\mathbf{E}\right\vert ^{2}$) of the incident field to a level that affects
the gaseous medium: ionization of the neutral atoms generates a plasma which in turn
has an effect of defocalization. The process toward singularity is stopped and a
balance between the two opposite tendencies become possible. A regime of
quasi-equilibrium is established with the structure of the beam in the
transversal plane becoming inhomogeneous: symmetrically centered on the axis
there is a channel of high beam intensity, having a diameter of the order of 
$100\ \mu $, surrounded by a much larger (diameter of the order of several
milimeters) cylindrical region where the intensity is much lower. In this
latter region most of the energy is located. This structure is a filament.
The fluctuation around the equilibrium, for powers in the beam higher than a
threshold $P_{in}>P_{cr}$ allows the propagation on large distances, that
can reach kilometers.

For powers that are a large factor (of the order of tens of units) greater
than the critical power $P_{cr}$ the picture changes. Essentially the
symmetrical structure of the propagation, mentioned above, becomes
azimuthally unstable \cite{skupinintense}. The azimuthal perturbations evolve under weak mutual
interaction and become sources of local filamentation, which dispose of
sufficient power ($>P_{cr}$) to propagate individually. In this
multi-filament structure the self-focalization followed by multiple
ionization and defocalization produced by the plasma take place for each
filament. The interaction leads to the random nucleation of filaments  \cite{nucleationrandom}, then
self-focusing followed by defocusing, everything leading to a random alternation of
filaments with short time of existence in a field of small amplitude where
the zones of plasma are dynamically redistributed in space \cite{Mlejnek98}, \cite{organizingmultiple}. This regime
has been called \emph{optical turbulence} \cite{mlejnekturbulence}. From measurements and
numerical simulations it results that the plane transversal to the beam
direction is organized in randomly meandering connected stripes (channels) where the
field has higher intensity (in this region there are the filaments too)
alternating and limited by, - similarly connected regions of plasma. This picture is clearly seen
in the Figures of the work by Ettoumi et al. \cite{phasetransition}.

\section{Hypothesis}

The qualitative aspect of the pictures suggests the following association:

\emph{the shape and the dynamics of the meandering structures (connected
ramnifications of two types) in the transversal plane to the direction of
propagation are identical with structures that result from the
reaction-diffusion dynamics of }$2D$\emph{\ media characterized by the
competition of two components: activator and inhibitor.}

This suggests to examine the possible common nature of the dynamics
associated with

\begin{itemize}
\item optical turbulence of the regime of multiple filamentation

\item the activator-inhibitor competition in nonlinear media
\end{itemize}

The simple qualitative comparison of the image of the transversal plane, and
respectively of the two-dimensional domain of a system activator-inhibitor
suggests that they may have a common nature (compare Figs. 1 and 2). Indeed , the activator-inhibitor
is universal and there would be no surprise to be found in particular
circumstances, as optical turbulence. The essential content of the two
phenomena is common: it is a competition between two components, with one
having auto-catalytic development and the other acting to limit and
eventually to suppress this effect.

\section{The equation of the envelope}

The propagation of the laser beam exhibits various regimes. For a power
greater than the critical threshold $P_{in}>P_{cr}$ the filament (central
channel and the surrounding energy bath) self-focuses up to the limit that
activates the opposing reaction of ionization followed by energization of
the electrons: multiphoton ionization and inverse brehmstrahlung energy
transfer. At much higher powers the same regime only persists for a finite
time (or length of propagation) followed by multi-filamentation and the
random dynamics produced by the modulational instability \cite{Couaironfilamentation}, \cite{bergephysrep}.

The electric field is represented in a multiple space-time analysis by
separating the slow evolution of the envelope $A\left( x,y,z,t\right) $, as $%
E=A\left( x,y,z,t\right) \chi \left( t\right) $ where $\chi \left( t\right)
=\exp \left( -i\omega _{0}t\right) $ is the fast wave factor. The nonlinear
polarization of the air, together with processes of interaction with the
plasma created by ionization, lead to the equation%
\begin{eqnarray*}
\frac{\partial A}{\partial z} &=&\frac{i}{2k_{0}}\Delta _{\perp }A-\frac{i}{2%
}\left( \frac{d^{2}k}{d\omega ^{2}}\right) \frac{\partial ^{2}A}{\partial
t^{2}}+i\omega n_{2}\left\vert A\right\vert ^{2}A \label{eqA} \\
&&-\frac{\sigma }{2}\rho A-\frac{i}{2}\sigma \omega \tau \rho A-\frac{\beta
^{\left( K\right) }}{2\sqrt{K}}\left\vert A\right\vert ^{2K-2}A  \nonumber
\end{eqnarray*}%
where $k_{0}$ is the central wavenumber of the beam and the terms represent:
the diffraction, the group-velocity dispersion (GDV), the Kerr nonlinearity
of the polarization, the transfer of energy from the beam to the electrons
of the plasma ($\sigma $ is the cross section of the inverse brehmstrahlung
effect) the rate of generation of the electrons ($\tau $ is the inverse of
the collision frequency) and the multi-photon ionization. A separate
equation, for the density of electrons $\rho $ is explained below. The equation above, which is a modified Non-linear Schrodinger 
equation, is integrated numerically in various regimes of beam power. For $%
P_{in}\gg P_{cr}$ the result shows that from a rather homogeneous
transversal state it occurs along propagation the nucleation of filaments.
The first to be lost is the azimuthal symmetry followed by the
quasi-independent evolution of the perturbations to definite filaments, by
concentration of the photon energy from the surrounding medium. The
self-focalization leads to episodic extinction and further re-nucleation of
filaments \cite{tzortzakis}. This is the regime of \emph{optical turbulence}. The equilibrium
becomes a dynamical state placed at marginal stability. There is a
competition between two opposite tendencies, and this is manifested as random
fluctuations in close proximity of a statistical equilibrium. For part of
the propagation, the fluctuations of the spatial distributions of the two
fields $A$ and $\rho $ have correlations that do not exhibit any intrinsic
scale, a situation that is characteristic to criticality. At longer
distances from the source, a sharp transition occur and the transversal
structure is broken into clusters of finite size \cite{phasetransition}.

\section{The common analytical structure of the optical turbulence and of
the \emph{activator-inhibitor} dynamics}

The transition from quasi-homogeneity in the transversal plane to a
structure consisting of connected stripes where the fluence is high separated by
similar branched channels of low fluence is similar to the fingering instability in a
reaction-diffusion system. In the latter case the interface separates two
distinct, competing, phases. In the high fluence region there is nucleation
of filaments. The analogous effect in the activator-inhibitor system is the
formation of \textquotedblleft spotty-spiky\textquotedblright\ solutions \cite{multibumpgm}.

The evolution of the envelope amplitude $A$ is associated with that of the
density $\rho $ of the electrons of the plasma generated by ionization at
focalization 
\[
\frac{\partial \rho }{\partial t}=D_{\rho }\Delta _{\perp }\rho +\frac{\beta
^{\left( K\right) }}{K\hslash \omega _{0}}\left\vert A\right\vert
^{2K}\left( 1-\frac{\rho }{\rho _{at}}\right) \label{eqrho}
\]%
Besides the last term that describes multi-photon ionization, we have
introduced a new term, absent in the standard treatments, of diffusion of
the electron density. We simplify the writing of the two equation
(introducing coefficients $a$, $b$, $\zeta $ and $\xi $) 
\begin{eqnarray*}
\frac{2k_{0}}{i}\frac{\partial \psi }{\partial z} &=&\Delta _{\perp }\psi
-a\left\vert \psi \right\vert ^{2}\psi +b\rho \psi \label{eqsima} \\
\frac{\partial \rho }{\partial t} &=&D_{\rho }\Delta _{\perp }\rho -\zeta
\rho \left\vert \psi \right\vert ^{2K}+\xi \left\vert \psi \right\vert ^{2K} \label{eqsimrho}
\end{eqnarray*}%
We first note that in the first equation above the term in the left and the first term in the right
side, if alone, would give a set of multiple (Gaussian-like) bumps, disposed
periodically on a line in the transversal plane. This may actually be seen
as a local limit of a circular contour where spots of high intensity (core
of filaments) exists (as confirmed by experiments \cite{mechain3}). This is very similar to what is found for
activator-inhibitor systems, where the spot solutions are also periodic \cite{weispotsgm}.

\begin{figure}[h]
\begin{minipage}{14pc}
\includegraphics[width=14pc]{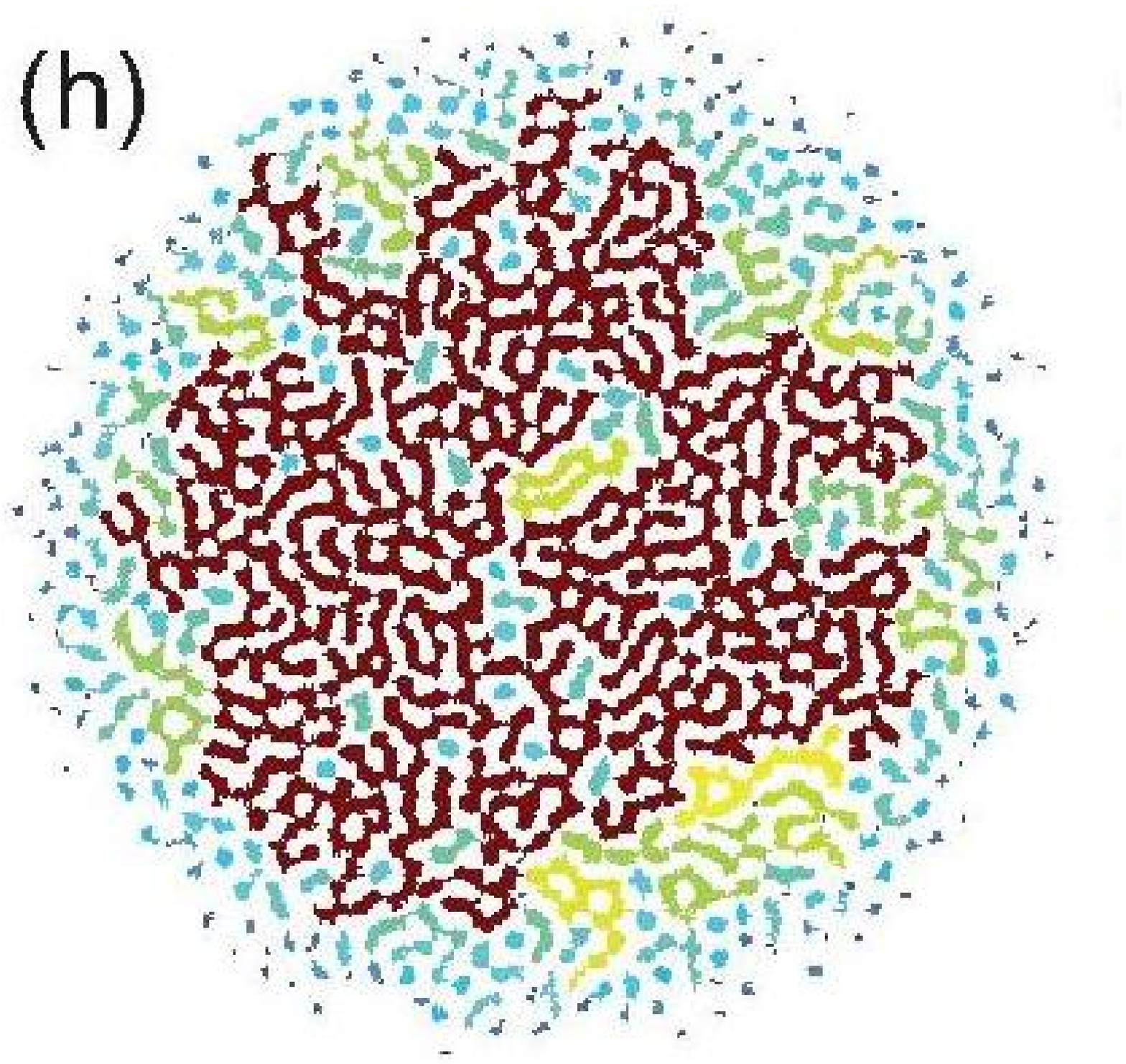}
\caption{\label{fig1}Structure of clusters in the transversal plane. This is subfigure (h) of Fig.1 from Ref \cite{phasetransition}. (Courtesy W. Ettoumi)}
\end{minipage}\hspace{2pc}%
\hfill
\begin{minipage}{14pc}
\includegraphics[width=14pc]{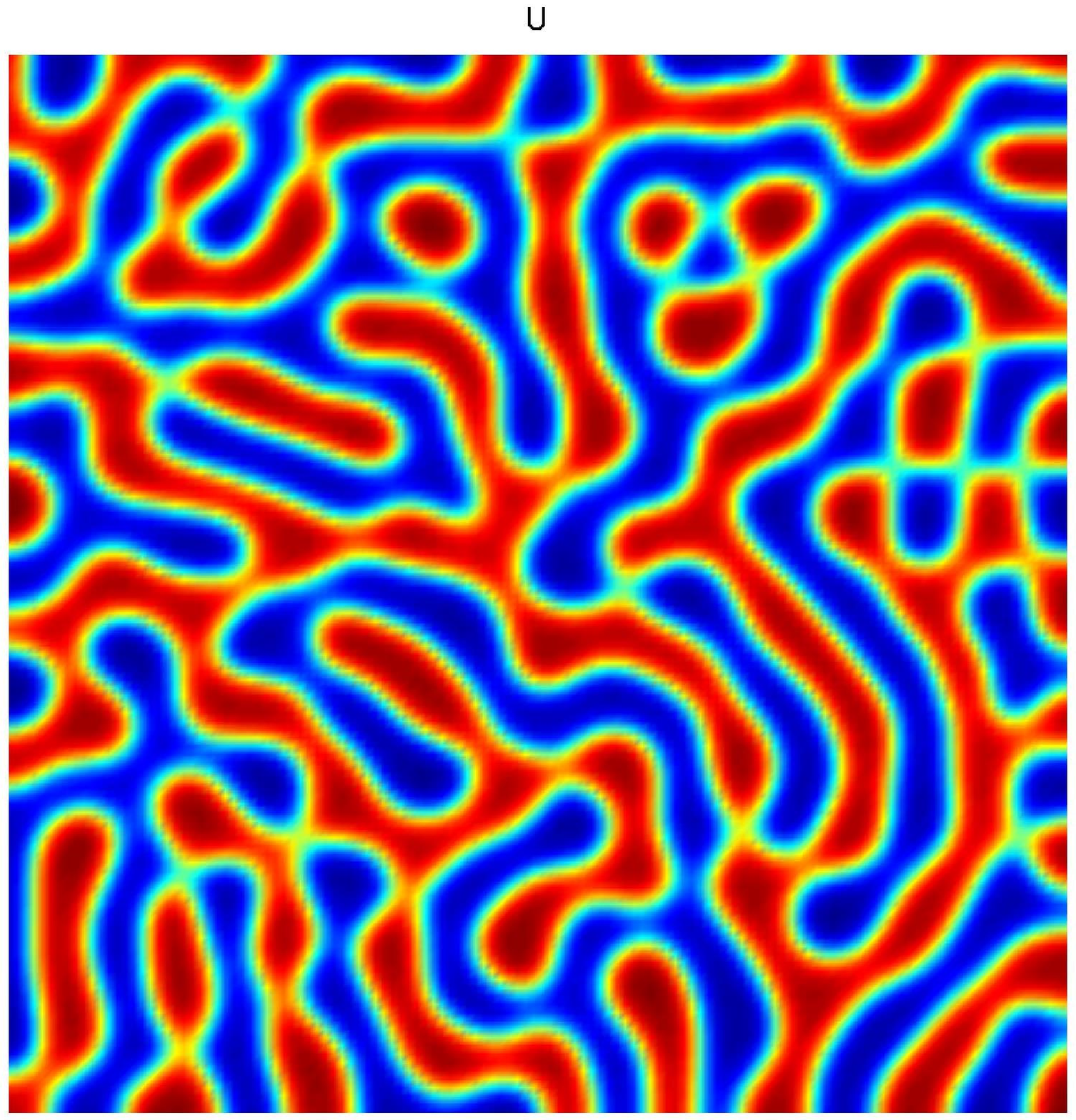}
\caption{\label{fig2}Labyrinth pattern for the Cahn-Hilliard activator-inhibitor system.}
\end{minipage} 
\end{figure}

We note the similarity between the system of equation from where it results
the \emph{optical turbulence} and the analytical structure of the model
FitzHugh-Nagumo, which exhibits the \emph{labyrinth instability} \cite{goldsteinlabyrinth}%
\begin{eqnarray*}
\frac{\partial u}{\partial t} &=&\varepsilon ^{2}\Delta u+f\left( u\right) -v
\label{fhn} \\
\frac{\partial v}{\partial t} &=&\Delta v-\delta \gamma v+\delta u  \nonumber
\end{eqnarray*}%
where $f\left( u\right) =u\left( 1-u\right) \left( u-u_{\kappa }\right) $
for $u_{\kappa }\in \left( 1,1/2\right) $. The regime in which this dynamics
consists of competing phases that occupy labyrinthic, mutually excluded,
connected channels (like clusters) needs a fast inhibitor $\left( v\right) $
diffusion.

The analogous behavior of the fields $\left( \psi ,\rho \right) $ and
respectively $\left( u,v\right) $ can be found to other models of the type 
\emph{activator-inhibitor}, like Gierer-Meinhardt, Gray-Scott, Cahn-Hilliard \cite{desaikapral}. The
picture of meandering ramnifications in plane is similar and has at the
origin the competition between the two physical fields \cite{goldsteinlabyrinth}, \cite{rolandrashimi}.

\bigskip

The activator-inhibitor systems have been shown to have a dynamics that is
constructed on a deep level of order, the \textquotedblleft shadow system\textquotedblright\ \cite{lini}. The existence of
this system is a common property of several reaction-diffusion models and
explains why they have similar behavior: labyrinthic interfaces, spot-like
solutions, their spatially regular distribution (polygonal), their
attraction and coalescence.

\section{Results that become accessible by the mapping between optical
turbulence and the \emph{activator-inhibitor} dynamics}

The analogy between the two systems may be useful. This is because the class
of \emph{activator-inhibitor} systems has been investigated mathematically
and disposes of precise description of its various solutions. We expect to
transfer some of these results to the model of optical turbulence,
especially for the regimes where it has been examined experimentally or by
numerical simulation

\begin{enumerate}
\item The solutions of the Gierer-Meinhardt (GM) system can be, for the case
of rapid diffusion of the \emph{inhibitor}, localized (\textquotedblleft
spot-like\textquotedblright ) \cite{multibumpgm}; this corresponds to the filaments observed in
the high-fluence region.

\item in some conditions, the GM solutions form groups (clusters) with
regular spatial disposition, eventually polygonal. Correspondingly, in a
laser beam it has already been found regular spacing of the multiple
filaments. It seems supported theoretically by the application of the notion
of \emph{Chaplygin gas with anomalous polytropic} \cite{trubnikovzhdanov}.

\item the solutions of GM, found to be grouped in clusters, have been proved
to be unstable and a number of \textquotedblleft spots\textquotedblright\
disappear after an evolution in time, being replaced by a single central
bump solution. This looks to be the analogous case to the coalescence of
filaments and re-formation of a single central filament.
\end{enumerate}

A problem raised by this mapping: which of the diffusion-reaction systems
that have a behavior of the type \emph{activator-inhibitor} can be
identified as the equivalent of the modified NSEq in the regime of
multiple-filamentation? The response appears to not be constraining, because at fast inhibitor 
they have the property of being manifestation of a \emph{shadow system}
which means a common type of behavior. However we must confine therefore to
those characteristics that are common and can be made to correspond to the
multi-filamentation.

A specific property is the proliferation of interfaces caused in the
activator-inhibitor system, by the \emph{labyrinth} instability and in the
laser field, by the competition between high fluence clustered (branched)
spatial regions (where filaments can nucleate as spots) and zones of plasma
with defocalization effect which keeps control on the local expansion of the
first phase.

An interesting aspect that can result from a comparative investigation
performed on the two systems is the effective interaction between filaments.
This is based on the connection between the NSEq (in its extended form for
beam propagation) and the Complex Ginzburg-Landau (CGL) equation \cite{nseglmalomed}. The exact
solution of the CGL equation is a soliton with an oscillating tail. If there
would be no interaction then the sum of two such functions would also be a
solution too. Replacing this sum in the expression of the energy, it should
result a sum of two times the individual energy of a single solution. Or,
this is not so, showing that besides the individual energies we have a term
of interaction. This depends parametrically on the positions of the centers
of the two solitons. When the relative distance between the centers is
varied, the supplementary term decreases (if there is attraction) or
increases (if there is repulsion). The method is identical with the one used
to find the interaction between vortices of the Abelian-Higgs
superconduction model \cite{JacobsRebbi} at non-self-duality. However
the energy of interaction between solitons of the CGL equation is found to
be exponentially small, which means that the coalescence of filaments is
slow. Or, the mapping to an \emph{activator-inhibitor} system helps to reformulate the problem: indeed there are solutions
consisting of several localized bumps with a space distribution which is
regularly periodic. This solution is unstable and the final state consists
of a central spike  \cite{cahnhilliard7}.

\bigskip

\section{Conclusion}

The filamentation generated during the propagation of the pulse of a high
power laser has many regimes and in particular the optical turbulence. It is
the formation in the transversal plane of a system of randomly meandering
ramnifications where the incident field is high, separated, and limited by -,  a
similar region dominated by defocusing plasma. This structure is dynamical
and in addition in the high intensity zone new filaments nucleate. They are
transient and end up by coalescing into a single chanel of propagation. This
regime can be mapped onto the activator-inhibitor dynamics of a nonlinear
reaction-diffusion system. Reformulated in the new framework, some problems
of beam propagation can be simpler. 

An important objective of further investigation is how is reflected in the
activator-inhibitor model the inverse phase transition that suppresses
progressively the long range correlations in the beam field, \emph{i.e.}
breaks up the large scale clusters.

{\bf Acknowledgments}

The authors thank to W. Ettoumi for permitting to reproduce a figure from the paper \cite{phasetransition}.

This work is partially supported by the Contract PN 09 39 01 01.


\begin{thebibliography}{10}
\expandafter\ifx\csname url\endcsname\relax
  \def\url#1{{\tt #1}}\fi
\expandafter\ifx\csname urlprefix\endcsname\relax\def\urlprefix{URL }\fi
\providecommand{\eprint}[2][]{\url{#2}}

\bibitem{Marburger1975}
Marburger J 1975 {\em Progress in Quantum Electronics\/} {\bf 4, Part 1} 35 --
  110 ISSN 0079-6727
  \urlprefix\url{http://www.sciencedirect.com/science/article/pii/0079672775900038}

\bibitem{Couaironfilamentation}
Couairon A and Mysyrowicz A 2007 {\em Physics Reports\/} {\bf 441} 47 -- 189
  ISSN 0370-1573
  \urlprefix\url{http://www.sciencedirect.com/science/article/pii/S037015730700021X}

\bibitem{bergephysrep}
Berg\'e L 1998 {\em Physics Reports\/} {\bf 303} 259 -- 370 ISSN 0370-1573
  \urlprefix\url{http://www.sciencedirect.com/science/article/pii/S0370157397000926}

\bibitem{skupinintense}
Skupin S, Peschel U, Etrich C, Leine L, Michaelis D and Lederer F 2002 {\em
  Opt. Lett.\/} {\bf 27} 1812--1814
  \urlprefix\url{http://ol.osa.org/abstract.cfm?URI=ol-27-20-1812}

\bibitem{nucleationrandom}
Kandidov V~P, Kosareva O~G, Tamarov M~P, Brodeur A and Chin S~L 1999 {\em
  Quantum Electronics\/} {\bf 29}(10) 911--915

\bibitem{Mlejnek98}
Mlejnek M, Wright E~M and Moloney J~V 1998 {\em Opt. Lett.\/} {\bf 23} 382--384
  \urlprefix\url{http://ol.osa.org/abstract.cfm?URI=ol-23-5-382}

\bibitem{organizingmultiple}
M\'echain G, Couairon A, Franco M, Prade B and Mysyrowicz A 2004 {\em Phys.
  Rev. Lett.\/} {\bf 93}(3) 035003
  \urlprefix\url{http://link.aps.org/doi/10.1103/PhysRevLett.93.035003}

\bibitem{mlejnekturbulence}
Mlejnek M, Kolesik M, Moloney J~V and Wright E~M 1999 {\em Phys. Rev. Lett.\/}
  {\bf 83}(15) 2938--2941
  \urlprefix\url{http://link.aps.org/doi/10.1103/PhysRevLett.83.2938}

\bibitem{phasetransition}
Ettoumi W, Kasparian J and Wolf J~P 2015 {\em Phys. Rev. Lett.\/} {\bf 114}(6)
  063903 \urlprefix\url{http://link.aps.org/doi/10.1103/PhysRevLett.114.063903}

\bibitem{tzortzakis}
Tzortzakis S, Berg\'e L, Couairon A, Franco M, Prade B and Mysyrowicz A 2001
  {\em Phys. Rev. Lett.\/} {\bf 86}(24) 5470--5473
  \urlprefix\url{http://link.aps.org/doi/10.1103/PhysRevLett.86.5470}

\bibitem{multibumpgm}
del Pino M, Kowalczyk M and Wei J 2003 {\em Ann. I. H. Poincar\'e\/} {\bf 20}
  53--85

\bibitem{mechain3}
M\'echain G, CD’Amico, Andr\'e Y~B, Tzortzakis S, Franco M, Prade B,
  Mysyrowicz A, Couairon A, Salmon E and Sauerbrey R 2005 {\em Optics
  Communications\/} {\bf 247} 171 -- 180 ISSN 0030-4018
  \urlprefix\url{http://www.sciencedirect.com/science/article/pii/S0030401804011812}

\bibitem{weispotsgm}
Wei J and Winter M 2008 {\em Journal of Mathematical Biology\/} {\bf 57} 53--89
  ISSN 0303-6812 \urlprefix\url{http://dx.doi.org/10.1007/s00285-007-0146-y}

\bibitem{goldsteinlabyrinth}
Goldstein R~E, Muraki D~J and Petrich D~M 1996 {\em Phys. Rev. E\/} {\bf 53}(4)
  3933--3957 \urlprefix\url{http://link.aps.org/doi/10.1103/PhysRevE.53.3933}

\bibitem{desaikapral}
Desai R~C and Kapral R 2009 {\em Dynamics of self-organized and self-assembled
  structures\/} (Cambridge University Press) ISBN 978-0-511-71966-0

\bibitem{rolandrashimi}
Roland C and Desai R~C 1990 {\em Phys. Rev. B\/} {\bf 42}(10) 6658--6669
  \urlprefix\url{http://link.aps.org/doi/10.1103/PhysRevB.42.6658}

\bibitem{lini}
Li F and Ni W~M 2009 {\em Journal of Differential Equations\/} {\bf 247} 1762
  -- 1776 ISSN 0022-0396
  \urlprefix\url{http://www.sciencedirect.com/science/article/pii/S0022039609001806}

\bibitem{trubnikovzhdanov}
Trubnikov B and Zhdanov S 1987 {\em Physics Reports\/} {\bf 155} 137 -- 230
  ISSN 0370-1573
  \urlprefix\url{http://www.sciencedirect.com/science/article/pii/0370157387901232}

\bibitem{nseglmalomed}
Malomed B~A 1991 {\em Phys. Rev. A\/} {\bf 44}(10) 6954--6957
  \urlprefix\url{http://link.aps.org/doi/10.1103/PhysRevA.44.6954}

\bibitem{JacobsRebbi}
Jacobs L and Rebbi C 1979 {\em Phys. Rev. B\/} {\bf 19}(9) 4486--4494
  \urlprefix\url{http://link.aps.org/doi/10.1103/PhysRevB.19.4486}

\bibitem{cahnhilliard7}
Alikakos N~D and Fusco G 1998 {\em Archive for Rational Mechanics and
  Analysis\/} {\bf 141} 1--61 ISSN 0003-9527
  \urlprefix\url{http://dx.doi.org/10.1007/s002050050072}

\end{thebibliography}
\providecommand{\newblock}{}

\end{document}